\documentclass[authoryear,nopreprintline]{elsarticle}

\usepackage{graphicx}
\usepackage{amssymb}

\journal{New Astronomy Reviews}

\begin{document}

\begin{frontmatter}

\title{The IAU Working Definition of an Exoplanet}

\author[inst1]{A.~Lecavelier des Etangs}

\affiliation[inst1]{organization={Institut d'Astrophysique de Paris, CNRS, UMR 7095, Sorbonne Universit\'e},
            addressline={98bis boulevard Arago}, 
            city={Paris},
            postcode={75014}, 
            country={France}}

\author[inst2]{Jack J. Lissauer\footnote{The two authors contributed equally to this manuscript and are listed in alphabetical order.}}

\affiliation[inst2]{organization={Space Science \& Astrobiology Division, MS 245-3, NASA-Ames Research Center},
            city={Moffett Field},
            postcode={94035}, 
            state={CA},
            country={USA}}
   
\begin{abstract}
In antiquity, all of the enduring celestial bodies that were seen to move relative to the background sky of stars were considered planets. 
During the Copernican revolution, this definition was altered to objects orbiting 
around the Sun, removing the Sun and Moon but adding the Earth to the list of known planets. 
The concept of planet is thus not simply a question of nature, origin, composition, mass or size, 
but historically a concept related to the motion of one body {\it around} the other, in a hierarchical 
configuration. \\
After discussion within the IAU Commission F2 ``Exoplanets and the Solar System'', the criterion of the star-planet mass ratio has been introduced in the definition
of the term ``exoplanet'', thereby requiring the hierarchical structure seen in our Solar System for an object to be referred to as an exoplanet. 
Additionally, the planetary mass objects orbiting brown dwarfs, provided they follow the mass ratio criterion, 
are now considered as exoplanets. 
Therefore, the current working definition of an exoplanet, as amended in August 2018 
by IAU Commission F2 ``Exoplanets and the Solar System'', reads as follows:\\
{\it Objects with true masses below the limiting mass for thermonuclear fusion of deuterium 
(currently calculated to be 13 Jupiter masses for objects of solar metallicity) 
that orbit stars, brown dwarfs or stellar remnants 
and that have a mass ratio with the central object below the $L_4$ / $L_5$ instability 
($M/M_{\rm central} < 2/(25+ \sqrt{621}) \approx 1/25$) are ``planets'', no matter how they formed. \\
The minimum mass/size required for an extrasolar object to be considered a planet should be the same as that used in our Solar System, which is a mass sufficient both for self-gravity to overcome rigid body forces and for clearing the neighborhood around the object's orbit.} \\
Here we discuss the history and the rationale behind this definition.
\end{abstract}


\end{frontmatter}

\newpage

\section{Introduction}
\label{Introduction}

\subsection{The importance of defining terms}

The definition of terms may appear to be unimportant for the advancement of science.
Whether or not the word ``planet" or ``exoplanet" is used for a newly discovered object 
is much less important for our knowledge than the understanding of its physical characteristics, 
its composition or its history and fate. Nonetheless, the primary purpose of language is to convey information in an efficient and effective manner. Usage of terms contributes to the definition in all living languages. This applies to scientific terminology,
where many words, even categories of objects like asteroids (see below), are defined more 
by usage than by a formal statement \citep{Wittgenstein_1953}. 

Scientific terminology is continuously improving 
because our collective knowledge is advancing. 
Refining the definition of terms contributes to the 
crystallization of knowledge. 
Clear and agreed-upon definitions facilitate communication of information.
In that perspective, the role of International Astronomical Union (IAU), which endeavors to promote global
collaboration, is crucial. 
One of the first achievements of the IAU, 
just after its birth in 1919, was the definition of the names of constellations  
and their boundaries on the sky \citep{delporte_1930}. 
Today, the IAU is tasked with adopting definitions of astronomical terms 
and names of astronomical objects, 
including those related to planets within our Solar System and beyond.

In 2001, the IAU's Working Group on Extrasolar Planets wrote a working definition for the term ``planet'' to be applied to extrasolar planets, which it subsequently amended in 2003. 
Fifteen years later, knowledge of exoplanets had significantly advanced, with  discoveries of a large number of planetary mass objects showing an astonishing diversity in physical properties. Therefore during 2018 we, as then President (ALdE) and Vice-President (JJL) of IAU Commission F2 - Exoplanets \& The Solar System, decided to lead our commission in a reassessment of the 2003 working definition.

\subsection{Historical Background -- Planets in the Solar System}
\label{sec:Planets in the Solar System}

The wonders of the night sky, the Moon and the Sun have fascinated mankind for many millennia. Ancient civilizations were particularly intrigued by several brilliant ``stars'' that move
among the far more numerous ``fixed'' (stationary)
stars. 
Aristotle and many other ancient Greek philosophers used and popularized the word 
$\pi \lambda \alpha \nu {\acute{\eta}} \tau \eta \varsigma$
(``{\it planete}''),  
meaning wandering star, to refer to these objects. 
The list of ``{\it planetes}'' included seven bodies in this category: 
the Sun, the Moon, Mercury, Venus, Mars, Jupiter and Saturn.

Although Aristarchus of Samos ($\sim$310 -- 230 BCE) had proposed the first known heliocentric model, the geocentric models of Aristotle and Ptolemy were more popular until the publication of the seminal work of \cite{Copernicus_1543}. Following the Copernican revolution, there were six bodies, all known to the ancients, that were classified as planets: Mercury, Venus, Earth, Mars, Jupiter and Saturn. This new classification give a similar list of 
planets with the exclusion of the Moon and the Sun and the addition of the Earth because it orbits the Sun as
do the other planets. The Copernican revolution changed the planet term from ``moving on the sky'', to ``moving around the Sun''. In both cases, motion was the key concept for the term ``planet'': the term 
planet was not related to the nature of the object given by its shape, size or composition, 
it was related to its dynamics. 

The first newly-discovered planet was Uranus, which was found by William Herschel in 1781. The discovery of Uranus was followed in rapid succession by the detection of four smaller objects traveling between the orbits of Mars and Jupiter: Ceres in 1801, then Pallas, Juno, and Vesta, the last of which was discovered in 1807. The total number of known objects classified as planets had reached 11, and remained at this value for almost four decades. A fifth small body between Mars and Jupiter, Astraea, was found in December of 1845, less than one year prior to the September 1846 discovery of Neptune. Three more small ``planets'' orbiting between Mars and Jupiter, now known as 6 Hebe, 7 Iris, 8 Flora, were detected in the second half of 1847, and several others were found in the next few years. It became clear that many objects orbited between Mars and Jupiter, and that all of them were much smaller than any of the six planets known to the ancients, as well as Uranus and Neptune. These small bodies were therefore reclassified as ``minor planets'' or ``asteroids'', and the number of ``planets'' dropped down to eight.
 
Pluto was found in 1930. Although it was much fainter than the other planets, no minor planets more distant than the Jupiter Trojans were known at the time, and Pluto was near the location predicted for a $\sim 6$ M$_\oplus$ (Earth mass) planet hypothesized to explain unaccounted for deviations of Neptune's orbit, so Pluto was quickly accepted as the ninth planet orbiting our Sun. 

However, subsequent data showed that Pluto was much smaller and less massive than initially thought. Pluto's absolute magnitude decreased as it approached perihelion, and further study showed that Neptune's orbit was better explained without a massive perturber at Pluto's location. Analysis of the orbit of Pluto's large moon Charon, which was discovered in 1978, implied that Pluto was only 4\% as massive as Mercury, the smallest of the eight other bodies regarded as planets.  

The minor planet 1992 QB1, later named 15760~Albion,
which like Pluto had a semimajor axis larger than that of Neptune, was discovered in 1992 \citep{Jewitt_1993}. 
Many other bodies were discovered in this region, referred to as the Kuiper Belt, in subsequent years. Pluto is by far the brightest Solar System body observed beyond Neptune's distance, but several known Kuiper Belt objects (KBOs) are larger than 1~Ceres, which is by far the largest asteroid. In 2005, the discovery of 136199~Eris, a KBO that is $\sim 25\%$ more massive than Pluto  \citep{Brown_2007} and that was initially thought to be slightly larger in size as well, brought some urgency to the question of how large a body needed to be in order to be classified a planet, because it didn't make sense to consider Pluto a planet if Eris was not considered a planet as well. 

A contentious discussion at the 2006 IAU General Assembly resulted in the adoption of the following definition for planets in the Solar System\footnote{Neither of the authors of this article was present for the 2006 planet definition and vote.}:

\noindent {\it A planet is a celestial body that:
\begin{enumerate}
\item is in orbit around the Sun, 
\item has sufficient mass for its self-gravity to overcome rigid body forces so that it assumes a hydrostatic equilibrium (nearly round) shape, and 
\item has cleared the neighborhood around its orbit.
\end{enumerate} }

\noindent Pluto was thus reclassified, and the number of Solar System planets dropped back down to eight.
 We note here that this definition is aimed at addressing the question 
of the use of the term ``planet'' for {\it objects in the Solar System}. 
Therefore, the first requirement to ``orbit around the Sun'' 
should not be read as an exclusion of exoplanets for the term
``planet'', but a restriction of the application of this definition to Solar System objects. 
 
\subsection{Historical Background -- Exoplanets}
 
At the beginning of this millennium, the IAU's Working Group on Extrasolar Planets (WGESP) wrote 
 and then amended a ``{\it working definition}" for the term ``planet" to be applied to exoplanets \citep{Boss_2003, Boss_2007}. 
 For more than 15 years, the definition was unchanged. It read as follows:
 
\noindent Position statement on the definition of a ``planet''

{\bf Working group on extrasolar planets (WGESP) of the International Astronomical Union}

Created: February 28, 2001

Last Modified: February 28, 2003\\

{\it Rather than try to construct a detailed definition of a planet which is designed to cover all future possibilities, the WGESP has agreed to restrict itself to developing a working definition applicable to the cases where there already are claimed detections, e.g., the radial velocity surveys of companions to (mostly) solar-type stars, and the imaging surveys for free-floating objects in young star clusters. As new claims are made in the future, the WGESP will weigh their individual merits and circumstances, and will try to fit the new objects into the WGESP definition of a ``planet'', revising this definition as necessary. This is a gradualist approach with an evolving definition, guided by the observations that will decide all in the end. 
Emphasizing again that this is only a working definition, subject to change as we learn more about the census of low-mass companions, the WGESP has agreed to the following statements: 
\begin{enumerate}
\item Objects with true masses below the limiting mass for thermonuclear fusion of deuterium (currently calculated to be 13 Jupiter masses for objects of solar metallicity) that orbit stars or stellar remnants are ``planets" (no matter how they formed). The minimum mass/size required for an extrasolar object to be considered a planet should be the same as that used in our Solar System.
\item Substellar objects with true masses above the limiting mass for thermonuclear fusion of deuterium are ``brown dwarfs", no matter how they formed nor where they are located.
\item Free-floating objects in young star clusters with masses below the limiting mass for thermonuclear fusion of deuterium are not ``planets", but are ``sub-brown dwarfs" (or whatever name is most appropriate).
\end{enumerate}
}

\noindent
{\it We can expect this definition to evolve as our knowledge improves.}

\section{Toward a new definition of an exoplanet}

\subsection{Need for an update}

As stated in the Introduction, 
the definition of a class of objects is a way to formalize our knowledge on these objects. 
By 2018, knowledge on exoplanets had significantly advanced since 2003. 
The number of known objects had been multiplied 
by a factor of 40 with the discovery of numerous exoplanets. These exoplanets have 
 a wide range of masses, radii and orbits, with many being quite different from any planets (in our Solar System or beyond), known in 2003. 
The diversity of system primaries also expanded, e.g., a 5~Jupiter mass object was  discovered 
in orbit around a 25~Jupiter mass brown dwarf \citep{Chauvin_2004}, 
raising new questions on the planethood of this kind of objects and others \citep[e.g.,][]{Basri_2006}.
So we considered it time for the 2003 working definition to be reassessed and, if needed, updated.

\subsection{The process}
\label{The process}

The Organizing Committee (OC) of IAU Commission F2 ``Exoplanets and the Solar System'' 
addressed the subject during the 2015--2018 triennium. 
After considering various possibilities, it was decided to consult the exoplanet 
community by questioning the whole membership of the commission. 
With about 400~members, this commission includes only a small 
subset of all astronomers working in the field. 
Nonetheless, it can be considered as a representative sample of experienced exoplanet researchers because of the wide diversity 
of its membership, both geographically  and regarding research activities 
(theory, observations, dynamics, characterization, etc.).

The Commission F2 Organizing Committee concluded that, while the definition should evolve with time, 
unless fundamental discoveries dictated otherwise, the evolution 
in the definition should be in continuity with the previous definition. 
As a consequence, it was decided that any change in the definition of an exoplanet 
should be made only when a substantial majority of the community 
considers it to be an improvement. To translate the principle of a ``substantial majority"
into numbers, a limit of a qualified majority of 2/3 was  agreed upon within the OC.  

Also, to make the decision practical and the process efficient, the possibility to ask general 
question to the community such as ``Should the definition (or some specific portion thereof) be changed'' 
has been excluded.
Indeed, a large majority can be in favor of some type of change, while no majority agrees on any specific change to be made. 
Therefore, specific amendments to the definition were prepared by the OC and proposed to a vote by 
the whole membership of the Commission.

Several issues on the mass of an exoplanet (lower and upper limits) were discussed within the OC, but only those supported by  a majority of the OC were
 proposed to the vote of the entire commission membership.
For instance, the OC agreed on introducing the planet-star mass ratio in the definition but 
did not agree on amending or discarding the 13 Jupiter mass limit,
so no amendment on this last issue was included in the plebiscite 
(\S\ref{The mass upper limit}).

\subsection{The lower limit}

Although the smallest extrasolar planets known to orbit 
actual stars (objects supported against gravitational collapse by self-sustained fusion, as opposed to stellar remnants like white dwarfs and neutron ``stars'')
are only $\sim 10^{-3}$ times 
as massive as any of the exoplanets known to orbit 
stars 
in 2003, none are in crossing orbits and all are 
significantly larger than Pluto and Eris. 
Also, perhaps more surprising, the planetary nature of PSR~1257+12b, with a mass of 1.8 times that of the Moon and orbiting a pulsar, has never been contested in the literature \citep{Schneider_2011}. Still, even for this extreme case, given the short orbital distance it can be shown that the criterion on the cleaning of the orbital neighbourhood is well satisfied \citep{Margot_2015}.    The same is the case for the Moon-size transiting exoplanet Kepler-37~b \citep{Barclay_2013}.

Thus, there has not been any significant controversy of the lower size/mass limit 
for any observed
object to be considered an exoplanet.
Given the lack of clear scientific direction from observations, 
the OC decided not to amend the wording of the 2003 provisional definition on the lower limit for an exoplanet, 
which explicitly refers to the criterion ``used in our Solar System''. Indeed, in 2006 the IAU formally adopted a definition for planets in the Solar System that specifically addresses the issue of the minimum mass (Sect.~\ref{sec:Planets in the Solar System}), 
and which can presently be used for extrasolar planetary systems.

The minimum planetary mass is therefore the mass sufficient for self-gravity to overcome rigid body forces and for clearing the neighborhood around the object's orbit. 
In other words, an object can be considered a planet if it is capable to influence dynamically the evolution of other bodies in its vicinity, sculpting their orbital distribution. 
This criterion has the advantage that the question of the evolution of other bodies in the neighborhood can be addressed through theoretical calculations or numerical simulations \citep[see, {\it e.g.},][]{Margot_2015}. \\
Note, however, that this definition of the lower limit relies on a dynamical criterion, which has been written with the architecture of the Solar System in mind. In future, it might need to be adapted to extrasolar systems showing different architectures, e.g.,with very eccentric orbits.

\subsection{The mass upper limit}
\label{The mass upper limit}

The upper mass limit of the 2003 definition has been controversial
 \citep{Chabrier_2014},
and various proposals have been made to alter it. 
Some researchers prefer definitions based on formation mechanism; although this has aesthetic appeal, it is highly impractical given the difficulty in determining the formation process of most bodies that might be reclassified by this change. 
The deuterium burning criterion should not be understood as a criterion linked to the formation mechanism.

Some researchers have proposed increasing the boundary to $\sim 25$ Jupiter masses (M$_{Jup}$) based upon the location of the ``driest'' region of the brown dwarf desert of the population of objects observed to orbit within a few AU of sunlike stars. However, there is no 
``gap'',  the location of this minimum appears to be a function of the mass of the stellar host, and it may also vary with orbital period. Furthermore, recent results 
show that the stellar metallicity - giant planet occurrence rate 
correlation goes mostly away for companion masses above 4~M$_{Jup}$ \citep[e.g.,][]{Schlaufman_2018}, 
which suggests it might be more appropriate to move the mass boundary downwards rather than upwards.
 
At present, the majority of known exoplanets have been discovered via transits, which observe size rather than mass. But a boundary based upon size would be inappropriate because the planetary radius function is not a monotonic function of  planetary mass, even for objects of a specified composition \citep{Stevenson_1976}.

As the Universe is less  than 14~billion years old, sizes of cool H/He-dominated objects increase with mass up to a peak 
at $\sim$4~M$_{Jup}$, and then drop (due to self-compression) to a minimum near 70~M$_{Jup}$, 
just below the start of the stellar main sequence. 
The radius of the star TRAPPIST-1 is very close to that of Jupiter, and 
even the Solar System's nearest stellar neighbor, Proxima Centauri, is smaller than many hot Jupiters.

Finally, 
the Organizing Committee decided to keep the current mass limit at the thermonuclear fusion of deuterium. 
This limit corresponds roughly to 13~Jupiter mass, with some dependence on the metallicity \citep{Spiegel_2011}. 

In conclusion, no amendment on this point was included in the plebiscite of the commission membership. 

\subsection{The mass ratio}

\begin{figure}[t]
  \centering
  \includegraphics[width=\columnwidth]{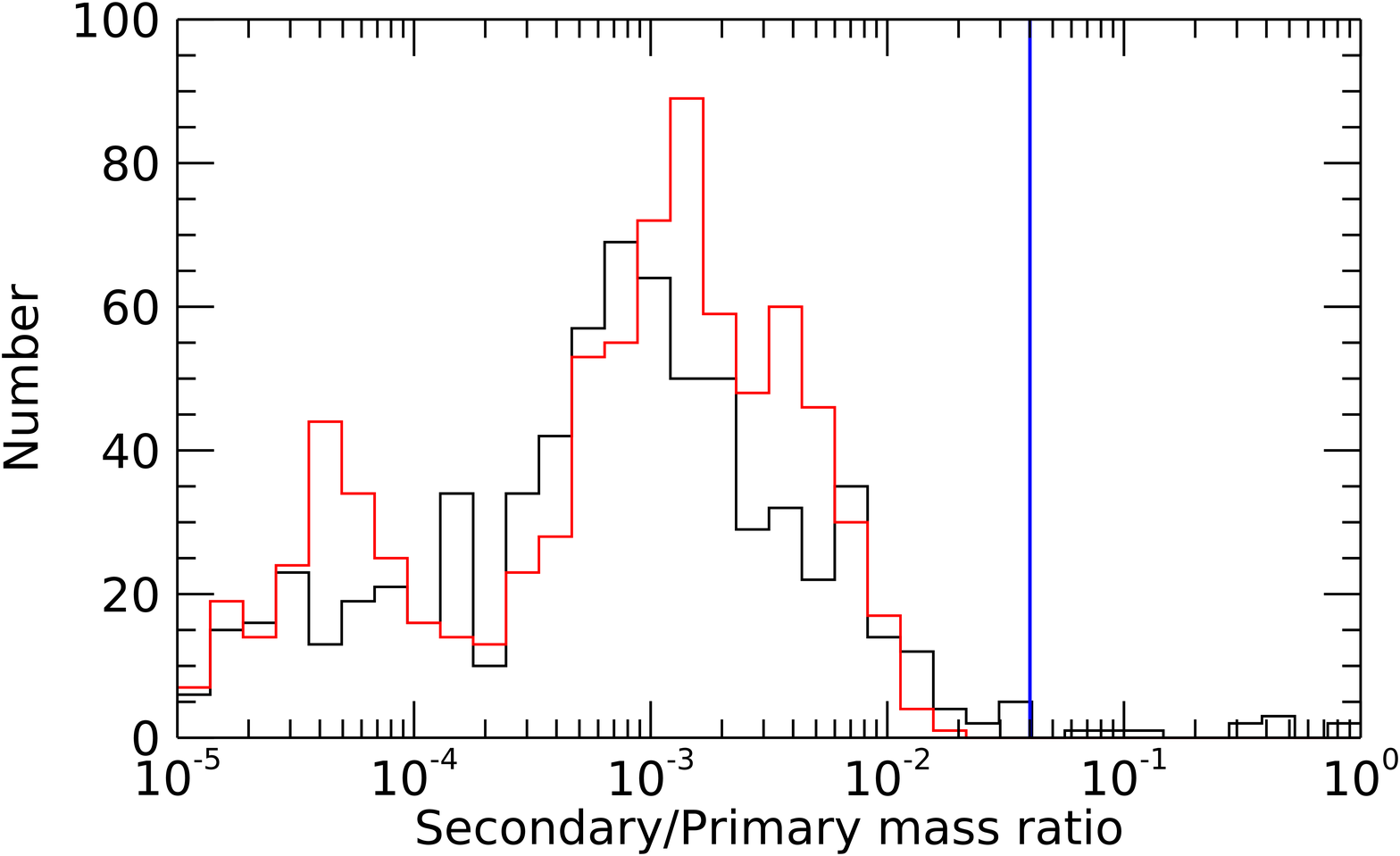}
   \caption{Histogram of the Secondary/Primary $q$ mass ratios for the known objects 
   with a mass of the secondary below 13 Jupiter mass. 
   The black histogram is for the measured mass,
   and the red histogram is for the $M\sin i$ measured using radial velocimetry. 
   The blue vertical line corresponds to the limit ratio of 1/25.
   The data are from The Extrasolar Planets Encyclopaedia accessed on September 16, 2021 
	 ({\tt exoplanet.eu}).
   }
     \label{fig:histogram}
  \end{figure}

\begin{figure}[b]
  \centering
  \includegraphics[width=\columnwidth]{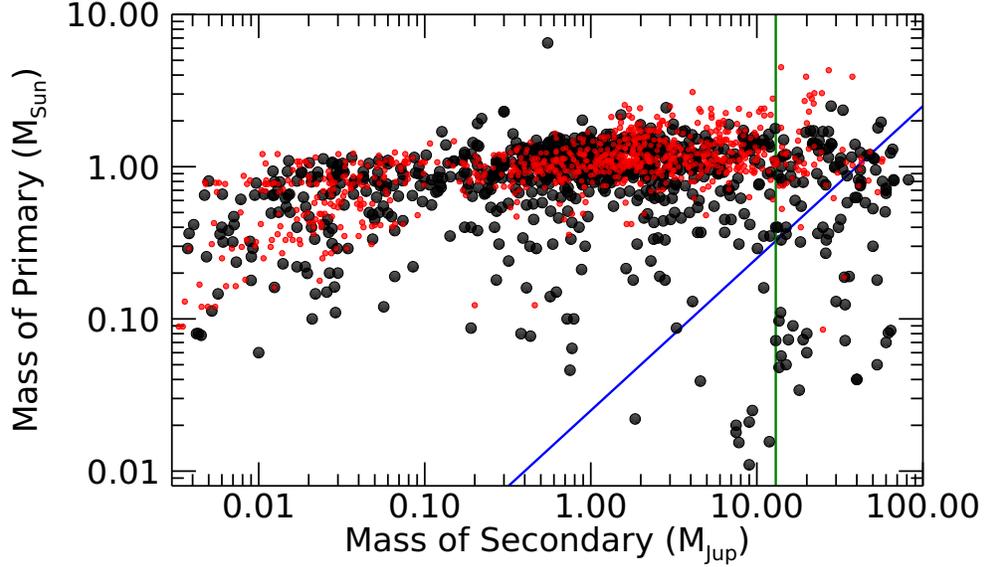}
   \caption{Primary-Secondary mass diagram. 
   The black dots are for the cases where the secondary mass has been measured, 
   and the red ones are for the cases where only the secondary $M\sin i$ has been measured 
   using radial velocimetry. 
   The green line corresponds to the 13 Jupiter mass limit. 
   The blue line corresponds to the Secondary/Primary mass ratio limit of 1/25.
   The data are from The Extrasolar Planets Encyclopaedia accessed on September 16, 2021 ({\tt exoplanet.eu}).
    }
     \label{fig:diagram}
  \end{figure}

At the time of the previous amendments to the exoplanet definition in 2003, no planetary mass objects had been found in orbit about brown dwarfs, and such objects were not considered in the 2003 definition. Members of this class have subsequently been found, and they have typically been referred to as exoplanets. Most planetary mass objects orbiting brown dwarfs seem to fall cleanly into one of two groups: 
(1) very massive objects, with masses of the same order as the object that they are bound to 
and (2) much lower mass objects. 
The high-mass group of companions appear akin to stellar binaries, whereas the much less massive 
bodies appear akin to planets orbiting stars. There is a wide separation in mass ratio 
between these two groupings Figures \ref{fig:histogram} and \ref{fig:diagram}, so any definition with a ratio between $\sim$1/100 and 1/10 
would provide similar results in the classification of known objects. 

It is noteworthy that
the limiting mass ratio for stability of the triangular Lagrangian points, 
$M/M_{\rm central} < 2/(25+ \sqrt{621})\approx 1/25$, falls in the middle of this range.
Moreover, this ratio is a limit based on dynamical grounds, which distinguish between star-planet 
couples where the star dominates and the planet can ``{\it clear the neighborhood around its orbit}'' 
(when the mass ratio is below 1/25), and pairs of objects where the more massive body 
does not dominate the dynamics to the extent that 
the less massive body can be to a good approximation considered to be orbiting about an immobile primary. 
Therefore, we proposed using this dynamically-based criterion as the dividing point.

The triangular Lagrangian points are potential energy maxima, but in the circular restricted three-body problem
the Coriolis force stabilizes them for the secondary to primary mass ratio ($m_2/m_1$) below~1/25, 
which is the case for all
known examples in the Solar System that are more massive than the
Pluto\index{Pluto!Pluto--Charon
system}--Charon\index{Charon!Pluto--Charon system} system. The
precise ratio required for linear stability of the
Lagrangian points $L_4$ and $L_5$ is $m_2/m_1 < 2/(25 + \sqrt{621}) \sim 1/25$ 
\citep[see][]{Danby_1988}.
If a particle at
$L_4$ or $L_5$ is perturbed slightly, it will start to {\itshape librate}
about these points (i.e., oscillate back and forth, without
circulating past the secondary).

From an observational point of view, mass ratios are commonly used in statistical studies 
of exoplanet discoveries by microlensing \citep[e.g.,][]{Suzuki_2016,Suzuki_2018} 
or by transit photometry \citep[e.g.,][]{Pascucci_2018}. 
It appears that the planet-to-star 
mass ratio is not only the quantity that is best measured in microlensing light curves analysis, 
but also it may be a more fundamental quantity in some aspects of planet formation than planet mass \citep{Pascucci_2018}.
It can be considered as a natural criterion to be used in the definition of the term exoplanet.

\subsection{The question of unbound planetary mass objects}

Motion relative to the ``fixed'' stars was the defining aspect of the ancient definition 
of the term ``planet''. 
Motion about the Sun became a requirement for planethood as a result of the adoption 
of the heliocentric model of the Solar System. Orbiting a star (or a similar object) 
is a natural extension of this requirement for exoplanets. We therefore agreed to keep
that requirement for an object to be considered as an exoplanet 
to be ``orbiting'' around a more massive object. 
The need to orbit around a star or an analogous massive object 
to be considered as a planet is effectively an extension of the Copernicus revolution. 

For planetary mass objects that do not orbit around a more massive central
object, the term ``sub-brown dwarf'' has not been adopted in the usage by the community; rather, 
these objects are 
often 
referred to as ``free floating planetary mass objects''. These two terms are nowadays considered as synonymous. 
An alternative to the rather ambiguous term ``free floating'', to specifically underline the presence or absence of a central object, 
would be to use the term ``unbound''. Note that neither of these terms accounts for the situation when the  planetary mass object is orbiting a companion whose mass is below the deuterium-burning limit and/or the minimum mass ratio for stability of the triangular Lagrangian points, but the advantages of brevity in terminology may well dictate that one of these terms is nonetheless optimal.

\subsection{The C.F2 consultation of July 2018}

Following the process of a qualified majority of 2/3 within the OC as defined in \S\ref{The process},
four questions were asked the whole membership of Commission F2 of the IAU.
Questions \#3 and \#4 were about some recommendations for exoplanet nomenclature
and the approval of the criterion of a qualified majority of 2/3 within the whole commission 
for future changes in the definition of an exoplanet, planetary nomenclature, and discovery criteria, respectively.

The first two questions in the plebescite asked to the commission were directly related to the new definition of an exoplanet: 
\begin{enumerate}
\item {\bf Question 1:} Should the term ``planets'' only apply to objects that have a mass ratio to the central object below the L4/L5 instability limit: \\
($M/M_{\rm central} < 2/(25+ \sqrt{621}) \approx 1/25$)?
\item {\bf Question 2:} In the WGESP definition, to be considered to be a ``planet'', the object needs to orbit a star or a stellar remnant. It is proposed to extend the criterion for the central object to include also the ``brown dwarfs''.

\end{enumerate}

Questions \#1 and \#2 got approved by more than 2/3 of the expressed opinions, with  71\% and 90\% of the vote, respectively.
Thus, both amendments were approved by  large majorities and
have been incorporated into the working definition given in Section~\ref{definition}. Question \#4 also passed, so future amendments to the definition will also require a two-thirds majority for passage.
  
\section{The new definition}

\subsection{The current IAU working definition of an exoplanet}
\label{definition}

The current working definition of an exoplanet is as follows:\\
{\it 
\begin{enumerate}
\item Objects with true masses below the limiting mass for thermonuclear fusion of deuterium (currently calculated to be 13 Jupiter masses for objects of solar metallicity) that orbit stars, brown dwarfs or stellar remnants and that have a mass ratio with the central object below the $L_4$/$L_5$ instability 
($M/M_{\rm central} < 2/(25+ \sqrt{621}) \approx 1/25$) are ``planets'' (no matter how they formed). The minimum mass/size required for an extrasolar object to be considered a planet should be the same as that used in our Solar System.
\item Substellar objects with true masses above the limiting mass for thermonuclear fusion of deuterium are ``brown dwarfs'', no matter how they formed nor where they are located.
\item Free-floating objects in young star clusters with masses below the limiting mass for thermonuclear fusion of deuterium are not ``planets'', but are ``sub-brown dwarfs'' (or whatever name is most appropriate).
\end{enumerate}
}

\subsection{What's new? A few examples}

The two changes between the new and the previous working definition of an exoplanet are the addition of the
$\sim$1/25 mass ratio criterion and the possibility to orbit a brown dwarf. 
These changes remove from the exoplanet list some massive objects, 
whose masses place them above the mass ratio criterion, 
and add new objects that orbit brown dwarfs. 
 
As a result, ten known systems with a mass ratio above 1/25 do not meet 
the criteria be considered as exoplanets, but as gravitationally linked 
binary systems composed of a sub-brown dwarf and a slightly more massive object. 
In one case, the host is a star, so the companion was classified as a planet under the previous definition, whereas in nine cases the host is a brown dwarf, so the secondary does not fulfil either the old or the new requirements for planethood. 
The characteristics of these binary systems are summarized in Table~\ref{tab:binary}.

In the other direction, five objects less massive than 13~Jupiter mass 
and orbiting around a brown dwarf are now considered to be exoplanets.
The characteristics of these exoplanetary systems are summarized in Table~\ref{tab:exoplanetary}.

\section{Future}
\label{Future}

There are some minor items that we think should be changed in the future. Near the beginning of item (3), the text ``in young star clusters'' is outdated, since similar, older objects have been seen outside of clusters, so it should be deleted. At the end of item (3), 
the text: ``sub-brown dwarfs'' (or whatever name is most appropriate) should be updated 
to the currently accepted term ``free-floating planetary mass objects''.

The exoplanet definition discussed herein is a working definition. 
The question of if/when it should go through a full IAU process 
to become the official IAU definition remains open. 
Indeed, in astronomy all definitions are to some extent ``working'', and even official ones 
are subject to amendments, because our knowledge and our understanding of the true nature 
of objects changes with time.

Note that the deuterium-burning limit is for solar metallicity. Strange objects like planetary mass bodies orbiting black widow pulsars \citep[e.g.,][]{Bailes_2011}, 
which probably lack hydrogen and are primarily composed of carbon and oxygen, 
are called planets because they are below 13~Jupiter masses and orbit a stellar remnant.

Both the WGESP and  IAU Commission F2 have avoided discussing the low mass limit because of the lack 
of data on exoplanets in the corresponding mass domain that precludes from addressing this issue. Very low mass ``exocomets'' have been observed to transit the star $\beta$ Pictoris since the mid-1980's, with detection of both the gas and the dust cometary tails
\citep{Kiefer_2014,Zieba_2019}. 
Small ``asteroids'' have been detected in very short-period orbits about white dwarfs 
\citep{Vanderburg_2015}, and ``disintegrating planets'' \citep[e.g.,][]{Croll_2014}
have been seen. 
The situation may change in the future when very low mass extrasolar orbiting objects will be discovered.
It appears that the current definition used for the Solar System, which defines the limit between
planets and dwarf planets, can be used for extrasolar objects, at least for those that clearly fall into one category or the other. When a significant number of intermediate extrasolar objects have been discovered, there
will be more information to address the possible need for new criteria in the definition. 

In conclusion, it must be emphasized that the current definition remains a working definition.
Current space missions like {\it Gaia}, {\it TESS} and {\it CHEOPS}, the missions to come like 
{\it PLATO} and {\it Nancy Grace Roman Space Telescope}, as well as ground-based observations, 
will increase the number of known objects significantly. 
As our knowledge of the exoplanetary zoo improves, the working definition 
might need to be amended or rewritten. \\

\section*{Acknowledgement}
The two authors contributed equally to this manuscript and are listed in alphabetical order.

We thank Antonella Barucci, Julio A. Fernandez, 
Ren\'e Heller, Antoniadou Kyriaki, Eric Mamajek,  
Rosemary Mardling, and Alessandro Morbidelli for constructive comments on preliminary drafts that enabled us to improve the manuscript.
We thank Cl\'ement Ranc for very useful discussions about the microlensing exoplanets. 

This research has made use of data obtained from the portal exoplanet.eu of The Extrasolar
Planets Encyclopaedia.

\bibliographystyle{elsarticle-harv}
\bibliography{bibliography}

\newpage

\newpage

\begin{table*}
\begin{center}
\begin{tabular}{lrrrr}
\hline
 Name  &  Primary & Secondary & M1/M2 & q=M2/M1 \\
       & mass & mass & Mass ratio & Mass ratio \\
       & ($M_{\rm Jup}$) & ($M_{\rm Jup}$) &  \\
\hline
MOA-2010-BLG-073L   & 160.  & 11.00 & 15.3  & 0.065 \\
KMT-2016-BLG-1820L  &  41.  &  4.57 &  8.8  & 0.113 \\   
OGLE-2011-BLG-0420L &  26.2 &  9.8  &  2.65 & 0.377 \\ 
OGLE-2012-BLG-0358L &  23   &  1.85 &  8.0  & 0.125 \\
MOA-2016-BLG-231L   &  21.0 &  9.00 &  2.33 & 0.430 \\
2M 0441+23          &  20.  &  7.50 &  2.67 & 0.370 \\
OGLE-2009-BLG-151L  &  18.9 &  7.9  &  2.39 & 0.419 \\
OGLE-2016-BLG-1266L &  15.7 & 12.0  &  1.31 & 0.762 \\
Oph 98              &  15.4 &  7.8  &  1.97 & 0.510 \\
WISE J1355-8258     &  11.  &  9.00 &  1.22 & 0.820 \\

\hline
\end{tabular}
\end{center}
\caption[]{Table of known binary systems including an object of less than 13~Jupiter mass 
and with a Secondary/Primary mass ratio above\,1/25.
The data are from the Extrasolar Planets Encyclop\ae dia \citep{Schneider_2011} accessed on September 16, 2021 
({\tt exoplanet.eu}).}
\label{tab:binary}
\end{table*}

\begin{table*}
\begin{center}
\begin{tabular}{lrrrr}
\hline
 Name  &  Primary & Secondary & M1/M2 & q=M2/M1 \\
       & mass & mass & Mass ratio & Mass ratio \\
       & ($M_{\rm Jup}$) & ($M_{\rm Earth}$) &  \\
\hline
 MOA-2007-BLG-192L   & 88. &   3.2 & 8300  & $1.2\times 10^{-4}$ \\
 OGLE-2015-BLG-1771L & 81. & 130.  & 200   & $5\times 10^{-3}$  \\         
 OGLE-2016-BLG-1195L & 80. &   1.4 & 17000 & $4.2\times 10^{-5}$  \\         
 KMT-2016-BLG-2605L  & 64. & 245.  & 83    & $1.2\times 10^{-2}$ \\ 
 OGLE-2017-BLG-1522L & 46. & 240.  & 63    & $1.6\times 10^{-2}$ \\        
\hline
\end{tabular}
\end{center}
\caption[]{Table of known exoplanetary systems where an exoplanet of less 
than 13~Jupiter mass orbit a brown dwarf.\\
For OGLE-2016-BLG-1195L the tabulated values are from \cite{Shvartzvald_2017} ; with a mass
of 0.2 $M_\odot$ estimated by \cite{Bond_2017}, the primary may not be a brown dwarf but a low mass star.}
\label{tab:exoplanetary}
\end{table*}

\end{document}